\documentclass{article}
\usepackage{amsfonts}

\usepackage{graphicx}
\usepackage{amsmath}

\begin{document}

\title{f-oscillators deformation for Moyal algebras}
\author{V.I. Man'ko$^a$, G. Marmo$^b$, E.C.G. Sudarshan$^c$, F. Zaccaria$^b$ \\
{\footnotesize \textit{$^a$P.N.Lebedev Physical Institute, Leninskii
Prospect 53, Moscow 119991, Russia }}\\
{\footnotesize {(e-mail: \texttt{manko@na.infn.it})}}\\
\textsl{{\footnotesize {$^b$Dipartimento di Scienze Fisiche dell'
Universit\`{a} ``Federico II" e Sezione INFN di Napoli,}}}\\
\textsl{{\footnotesize {Complesso Universitario di Monte S. Angelo, via
Cintia, 80126 Naples, Italy}}}\\
{\footnotesize {(e-mail: \texttt{marmo@na.infn.it, zaccaria@na.infn.it})}}\\
{\footnotesize \textit{$^c$Department of Physics, University of Texas,
Austin, Texas 78712, USA}}\\
{\footnotesize {(e-mail: \texttt{sudarshan@physics.utexas.edu})}}}
\maketitle

\begin{abstract}
Using general construction of star-product the q-deformed Wigner-Weyl-Moyal
quantization procedure is elaborated. The q-deformed Groenewold kernel
determining the product of quantum observables is given in explicit form for
small nonlinearities corresponding to nonlinear vibrations of classical and
quantum q-oscillators. The deformation of Groenewold kernel related to
general kinds of nonlinear vibrations described by f-oscillators are
considered. \noindent\textit{Key words} \newline
\end{abstract}

\section{Introduction}

By using the vector space of functions on phase-space one can
formulate both classical and quantum mechanics on the same carrier
space. What accounts for the difference between the two theories
is the product structure: it is associative, commutative and
pointwise for classical mechanics while it is associative,
noncommutative and nonlocal for quantum mechanics. This
formulation of quantum mechanics on phase space seems the most
appropriate to take care of the annoyance of Heisenberg as
expressed in a letter to Pauli \ \textquotedblleft \ldots the
worst thing is that I am quite unable to clarify the transition to
the classical theory\textquotedblright \cite {askBeppe1}.

It took almost twenty years before Moyal and Groenewold
independently arrived at a formulation of quantum mechanics on
phase space \cite{ Moyal, Grone}.

By using the Weyl correspondence associating operators to functions on phase
space \cite{Weyl} and the inverse one proposed by Wigner associating
functions to operators acting on some Hilbert space, many people have been
able to induce a star-product on functions by using the operator product
(see, e.g., \cite{Stratonovich, Zachos, Fronsdal, Berezin, Ba}). To be
precise, what Wigner did was to associate functions on phase space with
rank-one projectors (pure states). However by taking appropriate linear
combinations of these pure states it is not difficult to extend the
association to more general operators. In the same spirit one may consider
the diagonal representation introduced by Sudarshan \cite{Sud63} or the one
introduced by Husimi and Kano separately \cite{Husimi, Kano}.

The time evolution as given by the Moyal bracket is fully equivalent to the
one generated by the von Neumann equation for density states \cite{vonNeu}.
When the associative product is written in terms of integral kernels, the
commutative and associative product for classical mechanics has a kernel
expressed in terms of Dirac delta functions. The kernel appropriate \ for
the product of Weyl symbols corresponding to operators acting on some
Hilbert space was given by Groenewold and appear to be a twisted version of
the classical one: the twisting factor uses the symplectic structure
available on phase space. \ This circumstance makes clear that if we change
the symplectic structure by adding an electromagnetic field, the product of
function will change and therefore in general a deformation of the product
may be associated with the presence of an external field, i.e. it is a way
to incorporate such specific interactions. As the kernel is expressed in
terms of an exponent which uses the symplectic area and contains Planck's
constant as a parameter, when this parameter is set to zero one recovers the
kernel appropriate for classical mechanics \cite{Olga, Patrizia}.

This approach is often considered also to introduce twisted
products on space-time in the framework of noncommutative geometry
\cite{askBeppe2}. From what we have said, it should be clear that
any one-to-one correspondence between operators and functions will
induce a product on the space of functions, in particular one may
use symplectic tomographic probability distributions. The general
mathematical structure for the description of quantum states in
terms of probability distribution was clarified recently
\cite{Ventriglia1, Ventriglia2, Ventriglia3}. The operator-symbol
correspondence was also presented recently from a unified $pq$
anharmonic point of view by Klauder and Skagerstam \cite{Kla}.

Thus as the star-products are derived from the operator products, it is
clear that any deformation of the operator product, i.e. an alternative
product on the space of operators, will induce an alternative product in the
space of symbols. Alternative products in the space of operators were
considered by us in connection with a problem raised by Wigner \cite
{Wigner50} and analyzed to search bi-Hamiltonian descriptions at the quantum
level \cite{func(Si)}. An interpretation in terms of nonlinear phenomena
also with several physical consequences was given in \cite{f-osc}.

The $q$-oscillators \cite{Bid, Mc} were shown \cite{Salv} to correspond to
nonlinear vibrations of classical (and quantum) oscillators with specific
nonlinearity. The phase of such $q$-vibrations depends on the amplitude of
the vibrations quasi exponentially. In \cite{f-osc} the notion of $f$%
-oscillator was introduced to describe more general nonlinear vibrations in
the analogy of $q$-oscillators. The $f$-oscillators are deforming
commutation relations in very generic form which contains also the
particular case of the q-commutation relations corresponding to quantum
q-oscillators.

The $f$-oscillators describe the generic nonlinear relations in
which the frequency of vibrations depends on the energy of
vibrations and this dependence is determined by the function $f$.
The aim of the present work is to incorporate the $f$-nonlinearity
of the vibrations and to consider its influence on possible
deformations of the Moyal-Groenewold star-product. This product is
actually unique provided translational invariance is imposed as
 shown in \cite{SSZ}. Thus we study two kinds of deformations of point-wise product of
functions on classical phase space: the first deformation
associated with Planck constant providing the Moyal representation
of quantum mechanics while the next deformation uses extra
parameters (we will focus on q-deformation with extra
q-parameter). These deformations are associated with nonlinear
vibrations of $f$-oscillator and the nonlinear function $f$
dependent on energy of vibrations is exploited in our work as a
specific $K$-operator providing the $K$-deformation of associative
product of matrices in infinite dimensional Hilbert space. Another
goal of our work is to consider Lie algebra deformation based on
the deformed star-product of functions.

The paper is organized as follows. In section 2 we review the generic
star-product scheme and Moyal-Weyl-Wigner star-product scheme. In sect. 3 we
consider the $f$-oscillator formalism. In sect. 4 we construct the
q-deformed Moyal star-product and Moyal brackets. In sect. 5 the $f$%
-deformed Lie algebras will be discussed. The conclusions are
given in sect. 6.

\section{Weyl symbols and Moyal brackets}

In this section we review the general scheme of construction of
star-products on functions associated with operators \cite{Beppe1, Olga,
Patrizia, Patrizia1}.

This construction contains two main ingredients and we consider them at a
purely formal and algebraic level; in specific cases one has to check when
the appropriate conditions are satisfied.

Given a space $S$, one defines two families of operators thought of as
elements of a vector space $V$ and its dual space $V^{\prime }$, say for $%
x\in S$,\ $\hat{U}\left( x\right) \in V^{\prime }$ and $\hat{D}\left(
x\right) \in V$. \ We require that these two families allow the construction
of a partition of the unity for the space of operators we are interested in,
say
\begin{equation}
\mathbf{\hat{1}}=\int dx\hat{D}\left( x\right) \otimes \hat{U}\left( x\right)
\label{1}
\end{equation}
or
\begin{equation}
\mathbf{\hat{1}}=\sum_{n}\hat{D}\left( n\right) \otimes \hat{U}\left(
n\right)  \label{1'}
\end{equation}
when\ $S$\ is a discrete space.

With the help of these two families of operators we construct functions on $S
$, the symbols associated with operators by setting
\begin{equation}
f_{\hat{A}}(x)=\hat{U}\left( x\right) (\hat{A})  \label{2}
\end{equation}
and vice versa for any $f$ in the range of the previous map
\begin{equation}
\hat{A}_{f}=\int dxf\left( x\right) \hat{D}\left( x\right) \ .  \label{3}
\end{equation}
Clearly
\begin{equation}
\mathbf{\hat{1}\cdot }\hat{A}\mathbf{=}\hat{A}\mathbf{=}\int dx\hat{D}\left(
x\right) \hat{U}\left( x\right) (\hat{A})\ .  \label{4}
\end{equation}
Very often, when the space of operators we are considering contains inner
product, with abuse of notation we write
\begin{equation}
\hat{U}\left( x\right) \left( \hat{A}\right) =Tr\left( \hat{U}^{\dagger
}\left( x\right) \hat{A}\right) \ .  \label{5}
\end{equation}
The second ingredient consists of introducing a product structure on the
symbols in correspondence with any product on the vector space $V$. We set
\begin{equation}
\left( f_{\hat{A}}\ast f_{\hat{B}}\right) \left( x\right) =\int
dx_{1}dx_{2}f_{\hat{A}}\left( x_{1}\right) f_{\hat{B}}\left( x_{2}\right)
K\left( x_{1}.x_{2};x\right)   \label{6}
\end{equation}
where the integral kernel reads
\begin{equation}
K\left( x_{1}.x_{2};x\right) =\hat{U}\left( x\right) \left( \hat{D}\left(
x_{1}\right) \hat{D}\left( x_{2}\right) \right) =Tr\left( \hat{U}^{\dagger
}\left( x\right) \hat{D}\left( x_{1}\right) \hat{D}\left( x_{2}\right)
\right) \ .  \label{7}
\end{equation}
This kernel is a reproducing kernel and we have
\begin{equation}
\left( f_{\hat{A}}\ast f_{\hat{B}}\right) \left( x\right) =f_{\hat{A}\hat{B}%
}\left( x\right) \ .  \label{8}
\end{equation}
In the particular case of \ $S=\mathbb{R}^{2n}$ or $S=\mathbb{C}^{n}$ a
possible association is given by the Weyl correspondence
\begin{equation}
\hat{U}^{\dagger }(q,p)=2\pi \hat{D}(q,p)=2e^{\sqrt{2}[(q+ip)a^{\dagger
}-h.c.]}e^{i\pi a^{\dagger }a}  \label{9}
\end{equation}
where\ $a^{\dagger }$ and $a$ are bosonic creation and annihilation
operators (harmonic oscillator amplitudes) satisfying the commutation
relation
\begin{equation}
aa^{\dagger }-a^{\dagger }a=\mathbf{1}\ .  \label{10}
\end{equation}
One finds that, introducing complex coordinates, the operator $\hat{D}%
(\alpha ),$\ $\alpha \in \mathbb{C}$, $\alpha =\frac{1}{\sqrt{2}}(q+ip),$
becomes the displacement operator
\begin{equation}
\hat{T}(\alpha )=e^{\alpha a^{\dagger }-\alpha ^{\ast }a}  \label{11}
\end{equation}
giving rise to the unitary ray representation of $\mathbb{C}$%
\begin{equation}
\hat{T}(\alpha )\hat{T}(\beta )=\hat{T}(\alpha +\beta )e^{\frac{1}{2}(\alpha
\beta ^{\ast }-\alpha ^{\ast }\beta )}  \label{12}
\end{equation}
along with the trace formula
\begin{equation}
Tr\hat{D}(q,p)=\pi \delta (q)\delta (p)\ .  \label{13}
\end{equation}
It should be remarked that all bounded operators can be expressed
as a linear combination of such Weyl operators.
 The partition of unity requirement is satisfied in virtue of
\begin{equation}
Tr\hat{U}^{\dagger }(q,p)\hat{D}(q^{\prime },p^{\prime })=\delta
(q-q^{\prime })\delta (p-p^{\prime })\ .  \label{14}
\end{equation}
The Weyl symbol associated with the operator\ $\hat{A}$ is
\begin{equation}
f_{\hat{A}}(q,p)=2Tr[\hat{A}\hat{T}(2\alpha )e^{i\pi a^{\dagger }a}]\ .
\label{15}
\end{equation}
In the case of density state\ $\hat{\rho}$ eqs. \ref{14} and \ref{15} give
the expression for Wigner function of quantum state
\begin{equation}
W_{\rho }\left( q,p\right) =2Tr\left[ \hat{\rho}e^{\sqrt{2}\left(
q+ip\right) a^{\dagger }-h.c.}e^{i\pi a^{\dagger }a}\right] \ .  \label{16}
\end{equation}
The operator $e^{i\pi a^{\dagger }a}$ is the parity operator and using its
properties, e.g.
\begin{equation}
\hat{T}\left( \alpha \right) e^{i\pi a^{\dagger }a}=e^{i\pi a^{\dagger }a}%
\hat{T}\left( -\alpha \right)   \label{17}
\end{equation}
and eqs. \ref{10} and \ref{11} one can calculate the Groenewold kernel \cite
{Grone} for Weyl symbols
\begin{eqnarray}
K_{G}\left( q_{1},p_{1},q_{2},p_{2},q_{3},p_{3}\right)  &=&Tr\left[ \hat{D}%
\left( q_{1},p_{1}\right) \hat{D}\left( q_{2},p_{2}\right) \hat{U}^{\dagger
}\left( q_{3},p_{3}\right) \right]   \notag \\
&=&\pi ^{-2}e^{2i\left(
q_{3}p_{1}-q_{1}p_{3}+q_{1}p_{2}-q_{2}p_{1}+q_{2}p_{3}-q_{3}p_{2}\right) }\ .
\label{18}
\end{eqnarray}

\section{f- and q-oscillators}

The standard bosonic commutation relation, eq.(\ref{10}), can be deformed,
e.g. providing $q$-commutation relation \cite{Bid, Mc}. Such a deformation
is described \ by the generic formalism of $f-$oscillators \cite{f-osc}.
Considering classical vibrations of nonlinear oscillators \cite{Salv} it was
clarified \ that the dependence of phase of vibration on its energy provides
a generic deformation of the bosonic commutation relations. The deformation
is described by a function \ $f\left( a^{\dagger }a\right) $ determining new
annihilation operator
\begin{equation}
A=af\left( a^{\dagger }a\right) \ ,\ \ \ \ A^{\dagger }=f\left( a^{\dagger
}a\right) a^{\dagger }\ .  \label{19}
\end{equation}
For \ $f=1$ one has the standard annihilation operator. This deformation
replaces the commutator (\ref{10})\ by the commutator
\begin{equation}
AA^{\dagger }-A^{\dagger }A=af^{2}\left( a^{\dagger }a\right) a^{\dagger
}-a^{\dagger }af^{2}\left( a^{\dagger }a\right) \ .  \label{20}
\end{equation}
This commutator can be rewritten in the form
\begin{equation}
\left[ A,A^{\dagger }\right] =F\left( A^{\dagger }A\right)   \label{21}
\end{equation}
where the function $F$ is determined by the nonlinearity function\ $f\left(
a^{\dagger }a\right) $. For\ $f=1$\ we have $F=1$.

The $q$-oscillators are described by the specific function\
\begin{equation}
f_{q}\left( a^{\dagger }a\right) =\sqrt{\frac{\sinh (\lambda \ a^{\dagger }a)%
}{\lambda \ a^{\dagger }a}}\ ;\ q=e^{\lambda }\ .  \label{22}
\end{equation}
For $\lambda =0$ the $q$-parameter equals 1. For small parameter $\lambda $
which corresponds to weak nonlinearity of vibrations one has
\begin{equation}
f_{q}\left( a^{\dagger }a\right) \simeq 1+\frac{\lambda ^{2}}{12}\left(
a^{\dagger }a\right) ^{2}\ .  \label{23}
\end{equation}
The explicit form of nonlinearity of vibration follows from the equation of
motion for the amplitude of the oscillator. For standard oscillator with
unit frequency the equation reads
\begin{equation}
\dot{a}\left( t\right) =-ia\left( t\right)  \label{24}
\end{equation}
which has the solution
\begin{equation}
a\left( t\right) =ae^{-it}\ .  \label{25}
\end{equation}
One sees that the phase of vibrations does not depend on the amplitude. For
the nonlinear oscillator the equation of motion reads
\begin{equation}
\dot{a}\left( t\right) =-i\chi \left( a^{\dagger }a\right) a\left( t\right)
\ .  \label{26}
\end{equation}
Here the function\ $\chi \left( a^{\dagger }a\right) $ is a constant of the
motion and the solution to this equation reads
\begin{equation}
a\left( t\right) =ae^{-i\chi \left( a^{\dagger }a\right) t}\ .  \label{27}
\end{equation}
Thus the phase of the vibration depends on the energy of nonlinear
vibrations.\ Below we address the question how the nonlinearity of the
vibrations of the\ $f$-oscillator can be implemented to deform the Moyal
star-product. As we discussed the product is determined by quantizer and
dequantizer, which depend on the annihilation and creation operators$\ a$
and $a^{\dagger }$ of the standard harmonic oscillator. One can incorporate
the nonlinear vibrations using the nonlinearity function\ $f\left(
a^{\dagger }a\right) $. But if one simply replaces the oscillator operators $%
\left( a\rightarrow A=af\left( a^{\dagger }a\right) \right) $, in the
expressions for quantizer and dequantizer \ref{7} the compatibility
condition \ref{12} will be violated. We will then apply the function\ $%
f\left( a^{\dagger }a\right) $\ to deform the Moyal star-product using the
scheme of\ $K$-product which we below review.

\section{$K$-product of matrices}

In this section we review an approach to deform the product of matrices
keeping the associative property\cite{askBeppe3}. We consider such $K$%
-product in order to apply it to deform the Weyl-Moyal-Wigner product by
means of the nonlinearity function $\ f\left( a^{\dagger }a\right) $, used
in the previous section to construct $f$-oscillators.

The $K$-product is introduced by replacing the formula for associative
matrix product
\begin{equation}
c=a\cdot b\ \ \mathrm{or}\ \ c_{jk}=\sum\nolimits_{s}a_{js}b_{sk}\ ;\
j,k,s=1,2,\ldots ,N  \label{28}
\end{equation}
by the product
\begin{equation}
c=a\cdot _{K}b\ \ \mathrm{or}\ \
c_{jk}=\sum\nolimits_{s,m}a_{js}K_{sm}b_{mk}\ ;\ j,k,s,m=1,2,\ldots ,N
\label{29}
\end{equation}
where the matrix $K$ is used to define the $K$-product. Hereafter, $K$ may
depend on a parameter $\lambda$ in such a way that $K$ becomes the identity
when $\lambda$ goes to zero.

One can check that
\begin{equation}
\left( \left( a\cdot _{K}b\right) \cdot _{K}c\right) =\left( a\cdot
_{K}\left( b\cdot _{K}c\right) \right) \ .  \label{30}
\end{equation}
It means that the $\ K$-product is associative. For matrix \ $K=\mathbf{1}$
the $K$-product becomes the usual product of matrices.

However the new product has a unit only if $K$\ is invertible. Moreover if
we consider the one-to-one correspondence \ $A\rightarrow \sqrt{K}A\sqrt{K}%
\dot{=}\tilde{A}$ we find that there is a homomorphism\ $A\cdot B\rightarrow
\tilde{A}\cdot \tilde{B}$ which maps the identity to $K$. Of course the
square root exists only if $K$\ is positive$.$ For instance if $%
A_{1},A_{2},A_{3}$ close on the Lie algebra of the rotation group, the use
of a positive $K$\ will give rise to a new realization of the rotation
group. The matrix $K$ may be thought of as being responsible for replacing
the sphere, orbits of the standard realization of the rotation group, with
ellipsoids, orbits of the K-realization of the rotation group.

If one has matrices $a$ and $b$ corresponding to operators\ $\hat{a}$ and $%
\hat{b}$ acting in infinite dimensional Hilbert space the$\ K$-product of
the matrices is determined by the kernel of the product, i.e.
\begin{equation}
\left( a\cdot _{K}b\right) \left( x,x^{\prime }\right) =\int a\left(
x,y\right) K\left( y,z\right) b\left( z,x^{\prime }\right) dydz\ .
\label{31}
\end{equation}
The kernel\ $K\left( y,z\right) $\ determines the new $K$-product of the
operators acting on the Hilbert space.

\section{f-deformed Moyal product}

In this section we use the machinery of the $K$-product to deform the Moyal
product as was suggested in \cite{Patrizia2} and calculate in explicit form
the $q$-deformed Groenewold kernel for the deformation parameter $q$ close
to unity. By definition the deformation of associative product determined by
the nonlinear operator function $f\left( a^{\dagger }a\right) $ is given by
relation
\begin{equation}
f_{AB}\left( q,p\right) =Tr\left[ \hat{U}\left( q,p\right) \hat{A}f\left(
a^{\dagger }a\right) \hat{B}\right] \ .  \label{32}
\end{equation}
Using this definition and rewriting the symbol of product symbol in integral
form we get for the kernel of the product the relation
\begin{equation}
K_{f}\left( q_{1},p_{1},q_{2},p_{2},q,p\right) =Tr\left[ \hat{D}\left(
q_{1},p_{1}\right) f\left( a^{\dagger }a\right) \hat{D}\left(
q_{2},p_{2}\right) \hat{U}\left( q,p\right) \right] \ .  \label{33}
\end{equation}
The above kernel satisfies the properties of the solutions of associative
equation. To determine the kernel of $q$-deformed Groenewold kernel one
needs to calculate the following trace
\begin{eqnarray}
&&K_{\lambda }\left( q_{1},p_{1},q_{2},p_{2},q,p\right)   \label{34} \\
&=&\frac{2}{\pi ^{2}}Tr\left[ \left( 1+\frac{\lambda ^{2}}{12}\left(
a^{\dagger }a\right) ^{2}\right) e^{2\left( \alpha _{2}a^{\dagger }-\alpha
_{2}^{\ast }a\right) }e^{-2\left( \alpha a^{\dagger }-\alpha ^{\ast
}a\right) }e^{2\left( \alpha _{1}a^{\dagger }-\alpha _{1}^{\ast }a\right)
}\left( -1\right) ^{a^{\dagger }a}\right]   \notag
\end{eqnarray}
where
\begin{equation}
\alpha _{1,2}=\frac{1}{\sqrt{2}}(q_{1,2}+ip_{1,2})\ ,\ \alpha _{{}}=\frac{1}{%
\sqrt{2}}(q+ip)\ .  \label{35}
\end{equation}
The first term which does not contain the nonlinearity parameter $\lambda $\
coincides with Gronewold kernel
\begin{equation}
K_{G}\left( q_{1},p_{1},q_{2},p_{2},q,p\right) =\pi ^{-2}e^{2i\left(
qp_{1}-q_{1}p+q_{1}p_{2}-q_{2}p_{1}+q_{2}p-qp_{2}\right) }\ .  \label{36}
\end{equation}
One can calculate the nonlinear correction to the Groenewold kernel using
generating function for the deformed kernel.\ It means that we first
calculate the $f$-deformed kernel with the nonlinearity function of the form
\begin{equation}
f_{\tau }\left( a^{\dagger }a\right) =e^{i\tau a^{\dagger }a}\ .  \label{37}
\end{equation}
Having the $\tau $-deformed kernel we can get the correction in the\ $q$%
-deformed kernel by taking the second derivative of the $\tau $-deformed
kernel since
\begin{equation}
e^{-i\tau a^{\dagger }a}=1-i\tau a^{\dagger }a-\frac{1}{2}\tau ^{2}\left(
a^{\dagger }a\right) ^{2}+\ldots \ .  \label{38}
\end{equation}
Using the generating function we get the result
\begin{equation}
K_{\lambda }\left( q_{1},p_{1},q_{2},p_{2},q,p\right) =K_{G}\left(
q_{1},p_{1},q_{2},p_{2},q,p\right) \left[ 1+\frac{\lambda ^{2}}{192}\left(
\mu -1\right) ^{2}\right]   \label{39}
\end{equation}
where
\begin{equation}
\mu =\left( q-q_{2}-q_{1}\right) ^{2}+\left( p-p_{2}-p_{1}\right) ^{2}\ .
\label{40}
\end{equation}
The correction (\ref{40}) to the Groenewold kernel is symmetric with respect
to replacement
\begin{equation*}
q_{1}\longleftrightarrow q_{2}\ ,\ p_{1}\longleftrightarrow p_{2}\ .
\end{equation*}
It means that the Lie algebra structure constants related to Groenewold
kernel
\begin{equation}
C_{G}\left( q_{1},p_{1},q_{2},p_{2},q,p\right) \dot{=}K_{G}\left(
q_{1},p_{1},q_{2},p_{2},q,p\right) -K_{G}\left(
q_{2},p_{2},q_{1},p_{1},q,p\right)   \label{41}
\end{equation}
has the analogous deformed form
\begin{equation}
C_{\lambda }\left( q_{1},p_{1},q_{2},p_{2},q,p\right) =C_{G}\left(
q_{1},p_{1},q_{2},p_{2},q,p\right) \left( 1+\frac{\lambda ^{2}}{192}\left(
\mu -1\right) ^{2}\right) \ .  \label{42}
\end{equation}
The structure constants in eq.(\ref{41}) define infinite Lie algebra of Weyl
algebra. The structure constants in eq.(\ref{42}) define $q$-deformed on
finite Weyl algebra.

\section{Conclusions}

To conclude we point out some relevant aspects of the present work.

We realize that equations of motion on the algebra of operators
may start with a linear equation of the type $\frac{d}{dt}A=L(A)$
where $L$ is any linear map which does not need to respect the
product structure on the algebra. This is the case when we
consider generic Markovian evolution for open systems. When the
linear map turns out to be a derivation of the specific product we
are using, and the algebra is irreducible, the derivation
associated with $L$ will be an inner derivation and the dynamical
equation of motion becomes $\frac{d}{dt}A=\left[ H_{L},A\right] $,
i.e. acquires the Heisenberg form. Thus a given linear map $L$ may
be represented by alternative Hamiltonians according to the
alternative products it preserves: therefore the correspondence
``Hamiltonian $\longrightarrow $ equation of motion''\ depends on
the particular product we use. Therefore these three ingredients
appearing in the description of evolution may be used in different
manners. If we use a fixed Hamiltonian but change the product, we
get a different set of equations of motion (this would be the case
when the electromagnetic field is inserted in the Poisson
Brackets.) \ If we change the Hamiltonian \textit{along} with the
operator product we may\ \textit{compensate }the changes so that
the equations of motion would be the same (this occurs in the
description of biHamiltonian systems when dealing with complete
integrability). The particular instance we have considered,
changing the product while preserving the Hamiltonian has provided
us with an harmonic motion where the frequency depends on the
energy. It is thanks to this circumstance that the hydrogen atom
may be described as a reduction of harmonic oscillators
\cite{askBeppe4}. For the Harmonic oscillator the frequency and
phase are independent of amplitude (the action variable). So if
there are several harmonic oscillators with different amplitudes
they will keep their relative phases and all of them can be
brought to rest by a rotating phase space. On the other hand, for
the Kepler problem, Kepler's third law states that larger radii
means lesser frequency: $T^2\sim R^3$. The orbits of the nonlinear
oscillators can all be obtained by the central force which varies
in specified fashion on the action.

Thus, it is possible to consider `effective interactions' by changing the
product we use to multiply operators.

This point of view was the one taken, for instance, when $q$-deformed
oscillators were considered. Here we would like to stress that one may go
beyond these deformations and therefore go beyond Hopf algebras and their
deformations.

The other point of view we would like to stress is that by changing the
product structure on the space of operators we may obtain alternative
structures of $\mathbb{C}^{\ast }$-algebras on the same vector space of
operators; therefore the GNS construction appropriate for each $\mathbb{C}%
^{\ast }$-algebra structure would provide us with different Hilbert spaces,
which, due to the nonlinear change in the product structure, would be
connected in a one-to-one correspondence only if the nonlinear
transformations would be allowed. This becomes immediately clear if the
operators $A,A^{\dagger }$expressed by \ref{19} are used to construct the
Fock space out of the same vacuum as the one used by the usual $a,a^{\dagger
}.$

In conclusion, while this paper exhibits an explicit integral kernel for a
nonlinear deformation of the Moyal brackets, we seem to be suggested that
nonlinear transformations may find their way in the description of quantum
systems on Hilbert spaces by means of $\mathbb{C}^{\ast }$-algebras.

\section*{Acknowledgments}

V.I.M. and E.G.C.S. thank University of Naples "Federico II" for
kind hospitality. The work of V.I.M. was partially supported by
the Russian Foundation for Basic Research under Project No.
07-02-00598.

\end{document}